\documentclass{article}

\usepackage{arxiv}
\usepackage{cite}
\usepackage[utf8]{inputenc} 
\usepackage[T1]{fontenc}    
\usepackage{hyperref}       
\usepackage{url}            
\usepackage{booktabs}       
\usepackage{amsfonts}       
\usepackage{nicefrac}       
\usepackage{microtype}      
\usepackage{graphicx}
\usepackage[sort,compress,square,numbers]{natbib}
\usepackage{doi}
\usepackage[nottoc]{tocbibind}
\usepackage{float}
\usepackage{siunitx}
\usepackage{soul}
\usepackage{xcolor}
\usepackage[normalem]{ulem}

\title{Optimization of optical waveguide antennas \\for directive emission of light}

\author{ \href{https://orcid.org/0000-0001-7730-3489}{\includegraphics[scale=0.06]{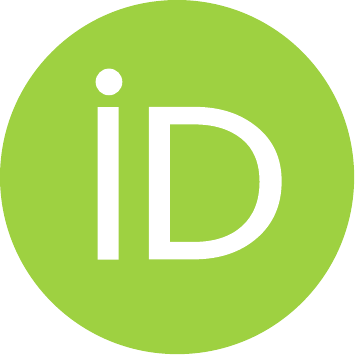}\hspace{1mm}Henna Farheen}\\
	Paderborn University\\
	Theoretical Electrical Engineering\\
	Warburger Str. 100, 33098 Paderborn, Germany \\
	\texttt{henna18@campus.uni-paderborn.de} \\
	\And
    \hspace{1mm}Till Leuteritz \\
	Universit\"at Bonn\\
	Physikalisches Institut\\
	53113 Bonn, Germany \\
	\texttt{leuteritz@physik.uni-bonn.de} \\
	\And
	\href{https://orcid.org/0000-0002-1028-9528}{\includegraphics[scale=0.06]{orcid.pdf}\hspace{1mm}Stefan Linden}\\
	Universit\"at Bonn\\
	Physikalisches Institut\\
	53113 Bonn, Germany \\
	\texttt{linden@physik.uni-bonn.de} \\
	\And
	\href{https://orcid.org/0000-0001-6431-746X}{\includegraphics[scale=0.06]{orcid.pdf}\hspace{1mm}Viktor Myroshnychenko} \\
    Paderborn University\\
	Theoretical Electrical Engineering\\
	Warburger Str. 100, 33098 Paderborn, Germany \\
	\texttt{viktor.myroshnychenko@uni-paderborn.de} \\
	\And
	\href{https://orcid.org/0000-0001-7059-9862}{\includegraphics[scale=0.06]{orcid.pdf}\hspace{1mm}Jens F\"orstner} \\
    Paderborn University\\
	Theoretical Electrical Engineering\\
	Warburger Str. 100, 33098 Paderborn, Germany \\
	\texttt{jens.foerstner@uni-paderborn.de} \\

}

\date{}

\renewcommand{\headeright}{}
\renewcommand{\undertitle}{}
\renewcommand{\shorttitle}{Optimization of optical waveguide antennas for directive emission of light}

\hypersetup{
pdftitle={Optimization of optical waveguide antennas for directive emission of light},
pdfkeywords={Directional emission, nanoantenna, dielectric, travelling-wave antenna},
}

\begin{document}
\maketitle

\begin{abstract}
Optical travelling wave antennas offer unique opportunities to control and selectively guide light into a specific direction which renders them as excellent candidates for optical communication and sensing. These applications require state of the art engineering to reach optimized functionalities such as high directivity and radiation efficiency, low side lobe level, broadband and tunable capabilities, and compact design. In this work we report on the numerical optimization of the directivity of optical travelling wave antennas made from low-loss dielectric materials using full-wave numerical simulations in conjunction with a particle swarm optimization algorithm. The antennas are composed of a reflector and a director deposited on a glass substrate and an emitter placed in the feed gap between them serves as an internal source of excitation. In particular, we analysed antennas with rectangular- and horn-shaped directors made of either Hafnium dioxide or Silicon. The optimized antennas produce highly directional emission due to the presence of two dominant guided TE modes in the director in addition to leaky modes. These guided modes dominate the far-field emission pattern and govern the direction of the main lobe emission which predominately originates from the end facet of the director. Our work also provides a comprehensive analysis of the modes, radiation patterns, parametric influences, and bandwidths of the antennas that highlights their robust nature.
\end{abstract}

\keywords{Directional emission \and Travelling-wave antenna \and Dielectric}

\section{Introduction}
Optical antennas have attracted much attention during the last two decades because of their interest from a fundamental viewpoint and practical importance \cite{bharadwaj2009optical}. Recent advances in the nanoscale fabrication enable scaling down the technology from the radio-wave regime to the optical regime and thus tailoring the light at nanoscale. This renders optical antennas as valuable building elements for new photonic structures and devices \cite{novotny2011antennas}. For instance, receiving optical nanoantennas have been widely investigated for improving the field enhancement and non-linear efficiencies \cite{bryant2008mapping,han2016high,gigli2019quasinormal}, while transmitting antennas have been rigorously studied for tuning and controlling the emission pattern by coupling to quantum emitters \cite{curto2010unidirectional,li2007shaping}.

In this proliferating field, directional optical antennas, which can radiate electromagnetic energy in a desirable direction, have been in the focus of photonic community due to their vast importance for many applications \cite{krasnok2013optical}. With regard to this ability, metallic nanoantennas which support surface plasmons benefit from a smaller footprint and resonant nature, but they also suffer from high ohmic losses due to absorption \cite{giannini2011plasmonic,halas2011plasmons,hoang2015ultrafast}. This makes their alternative, dielectrics, more attractive due to their reasonable bandwidth \cite{mongia1994dielectric}, low dissipative loss \cite{kuznetsov2016optically}, the ability for single photon emission \cite{au2019high} and near-unity radiation efficiency with highly directional radiation patterns \cite{krasnok2012all}. In particular, high-index dielectric antennas made of Silicon, Hafnium dioxide, Germanium or Gallium phosphide have attracted considerable interest \cite{barreda2019recent,sain2019nonlinear,cambiasso2017bridging,bidault2019dielectric}. Besides those, many hybrid  metal-dielectric antennas have also been studied for these purposes demonstrating promising advances \cite{morozov2018metal,livneh2016highly,rusak2014hybrid}. An extensive overview of this emerging research area including potential applications is given in Ref.~\cite{biagioni2012nanoantennas}.

Within this context, the traveling wave antennas \cite{TWA,constantine2005antenna,Antenny} operating at optical frequencies in which a traveling wave of electromagnetic oscillations is propagated along their guiding structure have received significant attention in the last years \cite{agio2013optical,li2021directional}. They typically produce an axially symmetrical shape in their radiation pattern and maintain an adequate directivity over a broad frequency range. Also, leaky-wave antennas have been emerging as a vital subset of the traveling wave antennas \cite{milligan2005modern,sutinjo2008radiation} that allow power to be leaked over the whole length of a non-resonant guiding structure by the virtue of leaky modes \cite{oliner2007leaky,hu2009understanding,peng1981guidance,oliner1981guidance,torner1990leaky,tamir1986varieties,peter2017directional,leuteritz2021dielectric}. The earliest illustration of such a structure was a rectangular waveguide with a continuous slit to its side \cite{hansen1940radiating}, followed by a plethora of recent advancements in such structures, successfully demonstrating the highly directive emission furnished by them \cite{jackson2012leaky,mohsen2018fundamental,jackson2013recent,jackson2008role,song2011silicon,peter2017directional}.

The highly directive emission can be also mediated by the propagation of guided modes along these waveguides in contrast to their aforementioned leaky counterparts. Attributing their capabilities to regulate the direction and angular distribution of optical radiation over a large spectral range, makes these antennas plausible candidates for robust wireless on-chip communication technologies and sensing \cite{yousefi2012waveguide}. Actually, these kind of dielectric structures can support both, leaky and guided-wave propagation, depending on the modes excited, and the resulting interference between them can lead to highly directed radiation patterns. In particular, our previous work demonstrates a complex interplay between leaky and guided modes in broadband waveguide-like optical nanoantennas made from low-loss dielectric materials that are able to shape the emission pattern of quantum dot emitters in the far-field with high directivity \cite{leuteritz2021dielectric}.

The electromagnetic response and overall functionality of nanoantennas strongly depend on structure geometry, its size and material, which enables the control of the modes excited, the frequency as well as the angular radiation patterns. A careful design and optimization of the nanoantennas is crucial to achieve optimal functionality. For this purpose, different methods and strategies have been extensively used, such as classical gradient descent and particle swarm methods, genetic and evolutionary optimization strategies, inverse design and deep learning approaches, or their combinations \cite{molesky2018inverse,ma2020deep,jiang2020deep,wiecha2019design,feichtner2012evolutionary,briones2018particle,robinson2002particle}.

In this work we use full-wave numerical simulations in conjunction with a particle swarm optimization (PSO) algorithm to design and optimize guided-wave directional optical antennas of high directivity. The antennas consist of a dielectric reflector and a director lying on a glass substrate, and a dipole emitter, placed in the feed gap between them, serves as an internal source of excitation. In particular, we analysed antennas with simple rectangular- and horn-shaped directors made of either HfO$_2$ or Si. Our analysis of the optimized antennas reveals the presence of leaky and guided modes contributing to the final near- and far-field patterns and explains their respective roles in achieving a high directive gain. Specifically, our simulations show that the light emission is predominately furnished/governed by two TE guided modes in the director and exhibit high gains in directivity up to 138 over a wide wavelength range. Furthermore, our systematic study quantifies the influence of every geometrical design parameter used in the optimization on the calculated directive gain and presents a comparison of the individual attributes of all optimized structures. Our findings demonstrate the uniqueness and robustness of these optical antennas, which renders them as excellent candidates for sensing applications and optical interconnects.

\section{Model, numerical methods and implementation}
In this work, we optimize the emission properties of travelling wave optical antennas composed of two dielectric elements, a director and a rectangular-shaped reflector, made of either Hafnium dioxide (HfO$_2$) or Silicon (Si) deposited on a SiO$_2$ glass substrate with refractive index $n_{\mathrm g}=1.52$. The director has either a simple rectangular- or horn-shape. The schematic of the numerical setup and orientation of the antennas with respect to the $xyz$ coordinates is illustrated in Fig.~\ref{fig::fig1}a and Fig.~\ref{fig::fig4}a. The antennas are oriented in the $xy$-plane so that the general direction of wave propagation and maximum radiation are primarily towards the positive $x$ direction. The positive $z$-axis is aligned normal to the substrate surface and points towards the substrate. A dipole emitter (solid red dot) with an emission wavelength of $780$\,nm serves as internal light source imitating the behavior of quantum dots. It is placed in the feed gap between the reflector and the director, $10$\,nm above the substrate at the origin of $y$-axis. We use a dipole with dipole moment oriented perpendicular to the antenna axis, i.e. along $y$-axis, that ensures the strongest coupling to the antenna modes, as demonstrated in our previous work \cite{leuteritz2021dielectric}. The light emission assisted by excitation of leaky and guided waves excited by the dipole emitter in the dielectric antenna is directed towards the glass substrate (depicted in red). It is collected over a solid angle $4\pi$\,steradian as a function of emission direction defined by the polar angle $\theta$ measured with respect to the optical axis ($z$-axis) and the azimuthal angle $\varphi$ measured with respect to the antenna axis ($x$-axis).

For optimization, we use full-wave numerical simulations based on the finite integration technique (FIT) for calculation of electromagnetic fields in conjunction with very powerful particle swarm optimization (PSO) algorithm which features heuristic global optimization \cite{CST,kennedy1995particle}. The algorithm optimizes the defined cost function by first providing an initial set of candidate solutions, also known as the particle swarm, which it iteratively improves by the particle movements in the search-space. This movement is based on the best known positions of the particle, which is constantly updated by the best local position of each particle. Thus, the swarm is iteratively guided to optimal solutions. As this work addresses high directive emissions, the optimization of geometrical design parameters are performed realising maximum directive gain (directivity), $D$, of the antenna as the cost function
\begin{equation}
D=\max D(\theta,\varphi)=\frac{4 \pi U(\theta,\varphi)}{\int_{0}^{2\pi}\int_{0}^{\pi} U(\theta,\varphi) \sin(\theta) d\theta d\varphi},
\label{eq::eq1}
\end{equation}
where $D(\theta,\varphi)$ is the directive gain and $U(\theta,\varphi)$ is the angular radiation intensity of the antenna in a given direction. The PSO algorithm is realized using a swarm size of 30 for 85 iterations to maximize the defined cost function and it was found converging to an optimized geometry after approximately 33 iterations, i.e. 1000 evaluations. Extending the accuracy of the solution to avoid being stuck in a local optimum, the obtained results are corroborated by solving the problem with the local trust-region optimization (TRO) which also converges within roughly 900 evaluations to the same results.

\section{Results and Discussion}
\subsection{Optimization of the \texorpdfstring{HfO$_2$}{HfO\_2} Rectangular-Shaped Antenna}
\begin{figure}[ht]
\centering\includegraphics[keepaspectratio,width=\textwidth]{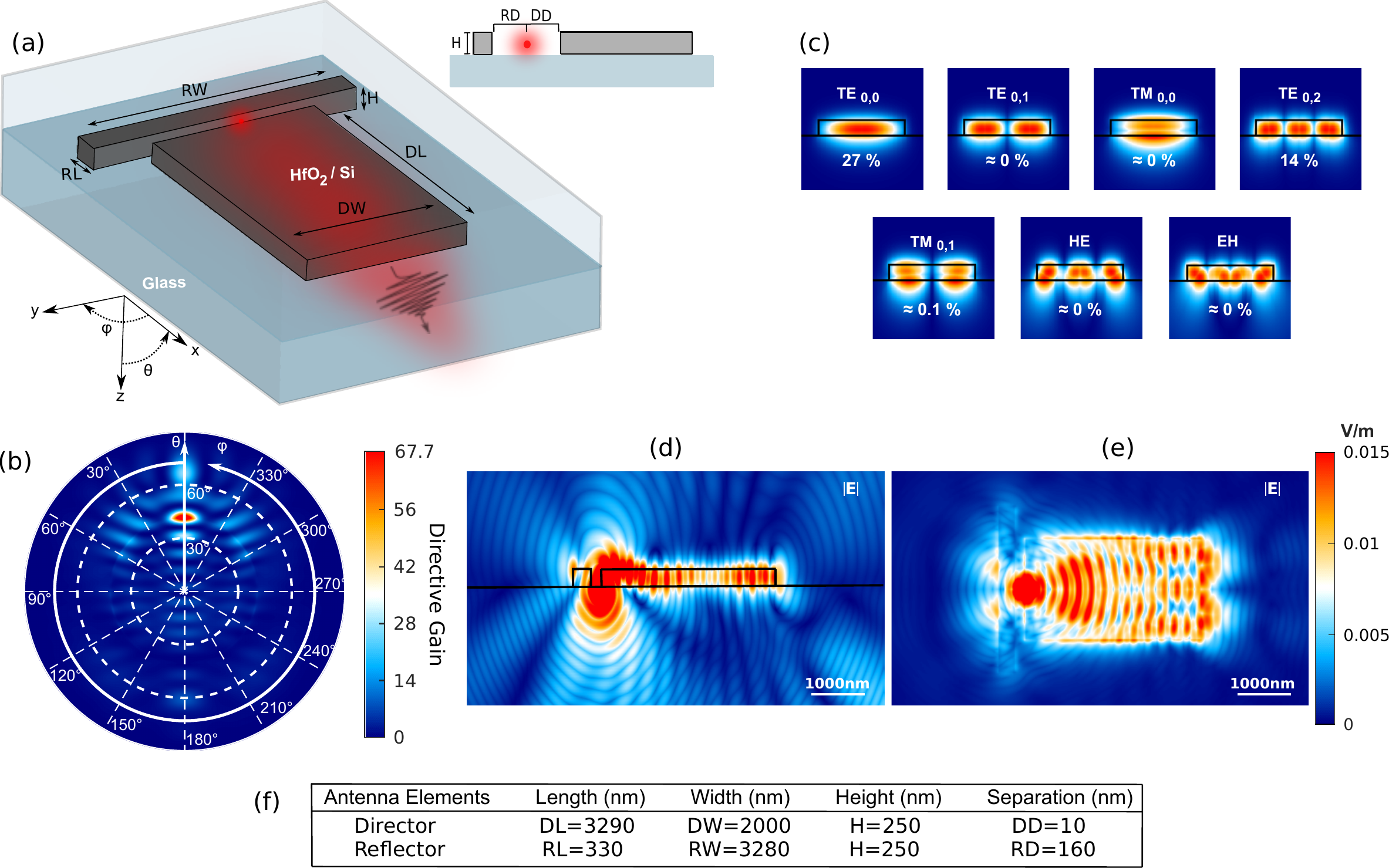}
\caption{Optimization of a HfO$_2$ rectangular-shaped antenna. (a) Schematic representation of our numerical setup. The antenna is composed of a HfO$_2$ rectangular-shaped director and reflector lying on a SiO$_2$ substrate and defined by seven design parameters affecting the directivity: the antenna height (H), director length (DL), director width (DW), reflector length (RL), reflector width (RW), distance of the field source from the director (DD) and the reflector (RD). The dipole source emitting at $780$\,nm with its dipole moment oriented along $y$-axis is placed into the feed gap between the director and reflector, $10$\,nm above substrate at the origin of $y$-axis (red dot). Light coupled to leaky and guided modes propagates along the director ($x$-axis) and emits radiation into the glass (depicted in red). (b) Calculated angular linear directive gain distribution of the optimized antenna exhibiting an in-plane directivity of $D=67.7$ at $\theta=41^{\circ}$ and $\varphi=0^{\circ}$. (c) The absolute electric field intensity distribution of the first seven guided modes excited by dipole emitter in the optimized director together with the amount of power coupled to them. (d,e) Calculated absolute value of the electric near-field $|E|$ (linear scale) of the optimized antenna in the (d) $xz$-plane at $y=0$ and (e) $xy$-plane at $z=0$, as produced by a $y$-oriented dipole emitter. (f) The table specifies the optimised design parameters of the antenna.}
\label{fig::fig1}
\end{figure}

We start our study by analysing a simple antenna consisting of a dielectric reflector and a rectangular-shaped director both made of HfO$_2$ with refractive index $n=1.9$ deposited on a SiO$_2$ substrate. In recent works, we have experimentally and numerically demonstrated that this kind of antenna is capable of high directivity of $29.2$ in magnitude \cite{peter2017directional,leuteritz2021dielectric}. Aiming to further improve the directivity, we firstly identify seven vital geometric parameters influencing the optimization problem: the antenna height (H), director length (DL), director width (DW), reflector length (RL), reflector width (RW), distance of the field source from the director (DD) and the reflector (RD). The schematic of the antenna highlighting the design parameters is illustrated in Fig.~\ref{fig::fig1}a.

Keeping directivity as the cost function, the optimization process converges to the optimized antenna geometry with design parameters given in table of Fig.~\ref{fig::fig1}. We note that our optimized antenna is much larger than the leaky-wave antenna demonstrated in Ref.~\cite{peter2017directional,leuteritz2021dielectric}. The directive gain emission pattern of the antenna resulted as a consequence of the interplay between the leaky and guided modes is depicted in Fig.~\ref{fig::fig1}b. It demonstrates superior directional properties with the main lobe emission directed into the substrate at $\theta=41^{\circ}$, in contrast to $\theta=72^{\circ}$ for the leaky-wave antenna. Moreover, this antenna shows a drastically improved linear directivity of $67.7$ compared to $29.2$ for its leaky-wave counterpart. These results suggest that the emission pattern of the optimized antenna is dominated by the traveling guided modes, though the weaker emission spot most likely caused by leaky modes is also observed at $\theta\approx68^{\circ}$. Furthermore, this antenna -- owing to the zero material-losses -- exhibits a near-unity radiation efficiency, which is defined as the ratio of the radiated power to the input power of the system.

In order to interpret the results and understand the origin of the radiation, we further analysed which modes are excited by the dipole in the antenna. We found that, in addition to the presence of the leaky-modes, the dipole couples to seven guided modes in the director which include three TE, two TM, and two hybrid modes with their mode-profiles shown in Fig.~\ref{fig::fig1}c, featuring the amount of power coupled to them. Notably, among them only the strongly excited TE$_{0,0}$ and TE$_{0,2}$ modes majorly influence the electromagnetic field with fractions of the total optical power coupled to them $27\%$ and $14\%$, respectively. The power coupled to each mode is extracted with the mode overlap product using the transverse near field components of each orthogonal mode. The rest of the power couples to the leaky modes of the antenna.

Indeed, the electric near-field along the structure shown in Fig.~\ref{fig::fig1}d exhibits a guided propagation along the director length, revealing the effects of consecutive multi-mode interference. The highest local electric field enhancement is induced in close proximity to the feeding area where the dipole power is efficiently coupled into the modes of the waveguide-like director. This is followed by the weakening of the field at the center of the director due to the destructive interference of the modes and, finally, the reinforced field at the end facet of the director is attributed to their constructive interference.

\begin{figure}[ht]
\centering\includegraphics[keepaspectratio,width=\textwidth]{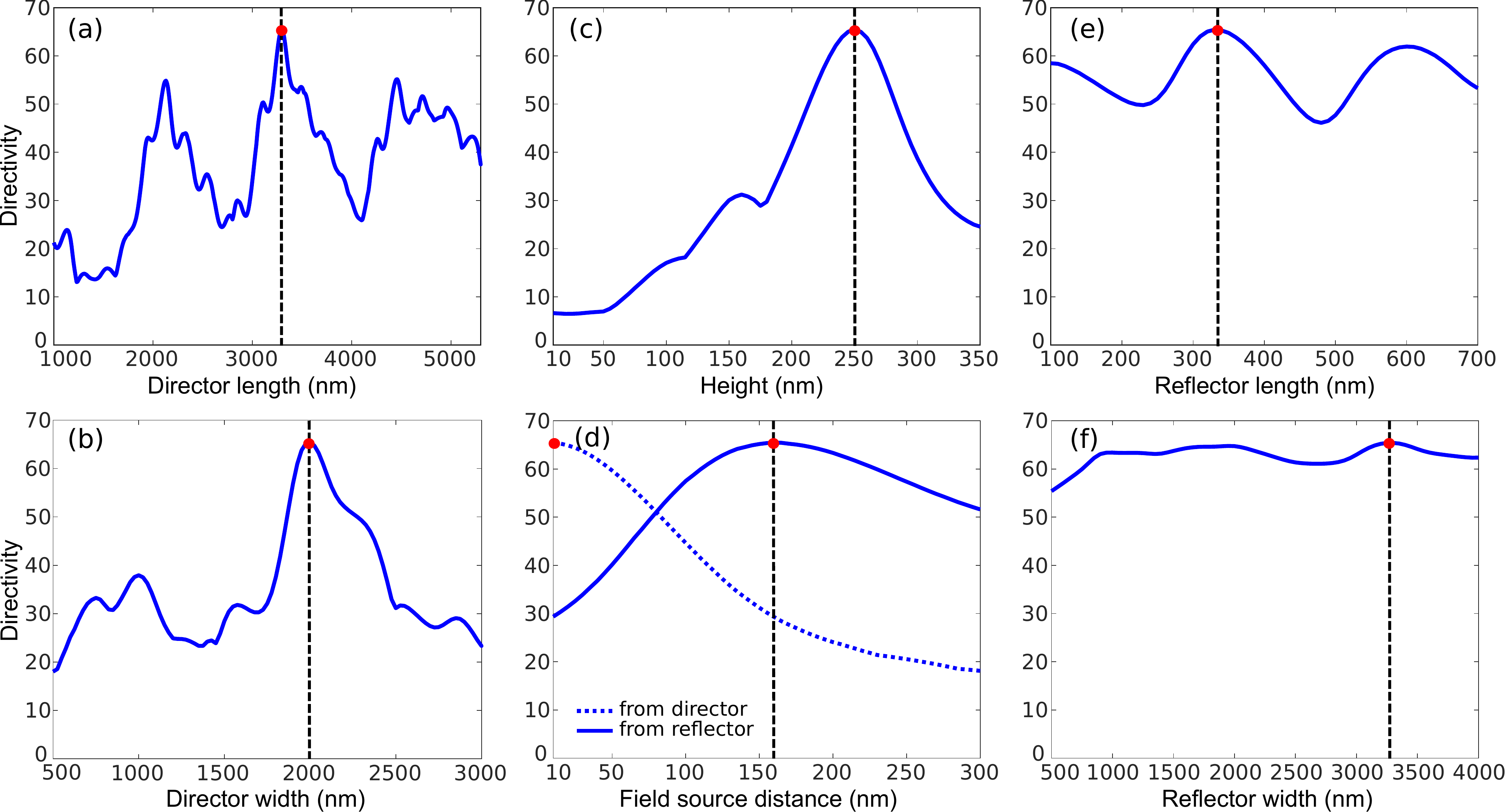}
\caption{Dependence of the directivity of the optimized HfO$_2$ antenna shown in Fig.~\ref{fig::fig1} on its seven design parameters: (a) the director length, (b) director width, (c) antenna height, (d) distance of the field source from the director and the reflector, (e) reflector length and (f) reflector width. The red point and dashed line on each plot represents the chosen optimal value of the corresponding parameter as shown in the table of Fig.~\ref{fig::fig1}, that results in the linear directivity of $67.7$.}
\label{fig::fig2}
\end{figure}

To check the robustness of the optimized antenna, we investigate the degree of influence of the design parameters on its directivity characteristics. As seen in Fig.~\ref{fig::fig2}, it is remarkable how the directivity can be significantly modulated by just regulating a few parameters. Here, the optima for all parameters are highlighted with red dots which together target the obtained highest directivity of $67.7$. For instance, Fig.~\ref{fig::fig2}a shows how the directivity rapidly changes as a function of the director length, which is constantly influenced by the state of interference fringes formed by the dominant guided modes at the end facets of these respective directors. Similarly, in Fig.~\ref{fig::fig2}b and c, the pronounced resonant peaks of directivity can be observed with ascending director width and antenna height, which are a consequence of the increased number of guided modes excited by the dipole source and their specific state of interference. In contrast, the directivity changes rather smoothly as a function of distance of the dipole source from the director and reflector (Fig.~\ref{fig::fig2}d). Specifically, the directivity of the system decreases as the distance of the dipole source from the director increases, caused by a weakening of the near-field coupling strength. Also, the optimal spacing between dipole source and reflector is found to be slightly lower than $1/4 \lambda$, typical for Yagi–Uda type antennas. Finally, the reflector length and width do not significantly influence the directivity as demonstrated in Fig.~\ref{fig::fig2}e and f, respectively, though the longer reflectors tend to exhibit a Fabry-P\'{e}rot behaviour.

\subsection{Optimization of the Si Rectangular-Shaped Antenna}
\begin{figure}[ht]
\centering\includegraphics[keepaspectratio,width=\textwidth]{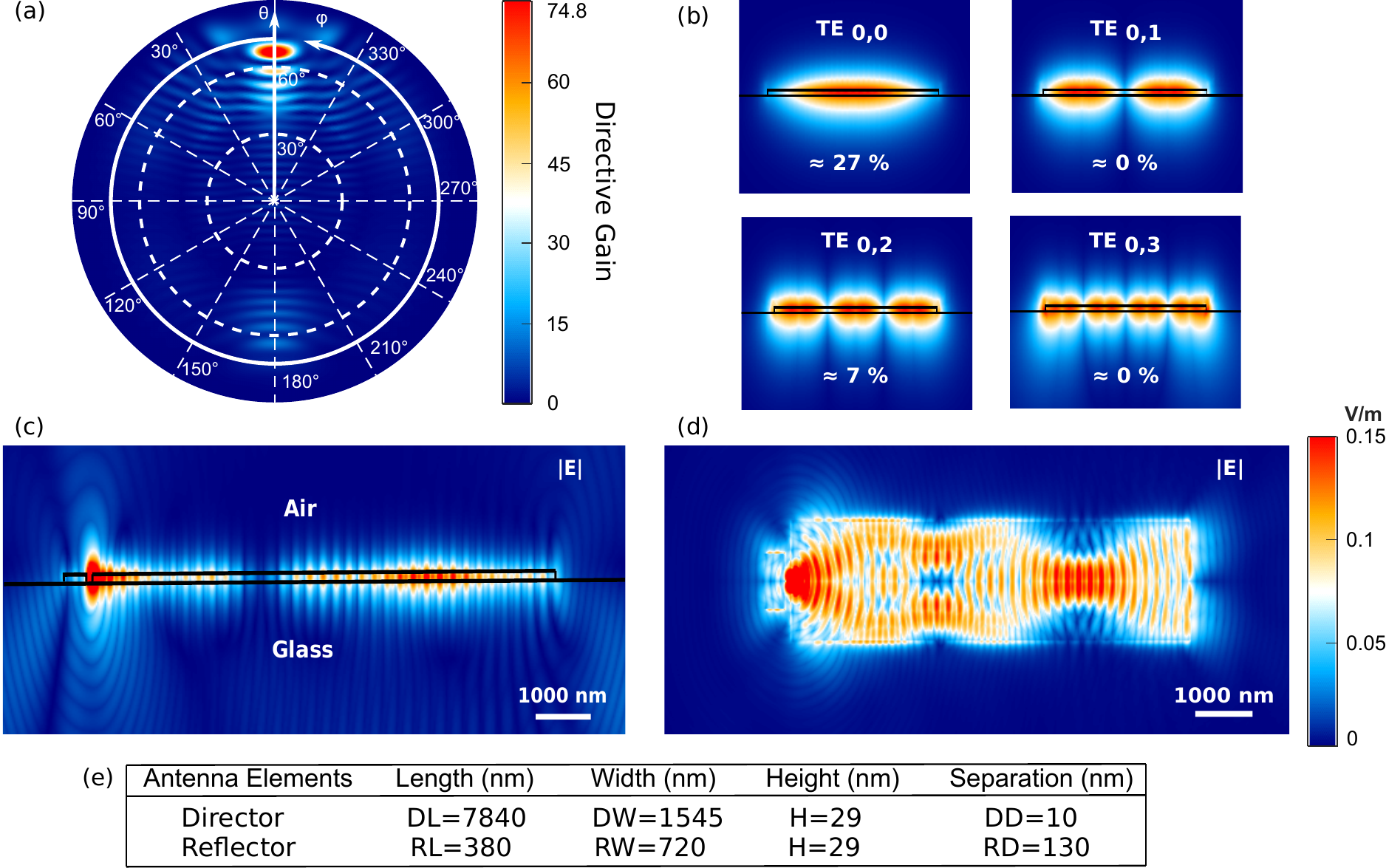}
\caption{Optimization of a Si rectangular-shaped antenna. Schematic representation of our numerical setup, the principle of operation, and design parameters used are demonstrated and explained in Fig.~\ref{fig::fig1}. (a) Calculated angular linear directive gain distribution of the optimized antenna exhibiting an in-plane directivity of $D=74.8$ at $\theta=66^{\circ}$ and $\varphi=0^{\circ}$. (b) The absolute electric field intensity distribution of four guided TE modes excited by dipole emitter in the optimized director together with the amount of power coupled to them. (c,d) Calculated absolute value of the electric near-field $|E|$ (linear scale) of the optimized antenna in the (c) $xz$-plane at $y=0$ and (d) $xy$-plane at $z=0$, as produced by a $y$-oriented dipole emitter. (e) The table specifies the optimised design parameters of the antenna.}
\label{fig::fig3}
\end{figure}

Being motivated by the extensive use of silicon in photonic devices in the last decade due to its extraordinarily rich linear and nonlinear optical properties \cite{sun2013large,lechago2019all,staude2017metamaterial,kivshar2018all}, we also optimized Si rectangular-shaped optical antenna with the same optimization setup and design parameters as in Fig.~\ref{fig::fig1}. Due to the large refractive-index contrast between Si ($n=3.71+0.007i$) and SiO$_2$, we expect to enhance the propagation characteristics of the antenna and thus attain the improved directivity. Keeping the same cost function, the PSO algorithm converges to the design parameters for the Si antenna given in table of Fig.~\ref{fig::fig3}. The resulting design retains the attractive directional properties of the HfO$_2$ antenna with the pronounced and focused main lobe and slightly increased side lobe level as seen in Fig.~\ref{fig::fig3}a. The antenna demonstrates a linear directivity of $74.8$ with the main lobe pointing at $\theta=66^{\circ}$ ($\varphi=0^{\circ}$), which are both higher than that of the HfO$_2$ antenna. Remarkably, due to the larger refractive index of Si, the director has a much smaller height in comparison to its HfO$_2$ counterpart to accommodate guided modes of approximately the same effective index like the ones present in the HfO$_2$ director \cite{leuteritz2021dielectric}. Indeed, the dipole source excites only four TE modes prohibiting any TM mode but the optical power dominantly couples again only to the two TE$_{0,0}$ and TE$_{0,2}$ guiding modes, with corresponding mode profiles illustrated in Fig.~\ref{fig::fig3}b. In particular, Fig.~\ref{fig::fig3}c and d nicely illustrate the guided but less-confined propagation of the waves and the periodic nature of their constructive-destructive interference along the director. We also note that this Si antenna constituting the material losses still shows a high radiation efficiency of approximately $94\%$. Dependence of the Si antenna directivity on its design parameters demonstrates a similar behaviour like its HfO$_2$ counterpart.

\subsection{Optimization of the \texorpdfstring{HfO$_2$}{HfO\_2} and Si Horn-Shaped Antennas}
\begin{figure}[h!]
\centering\includegraphics[scale=0.45]{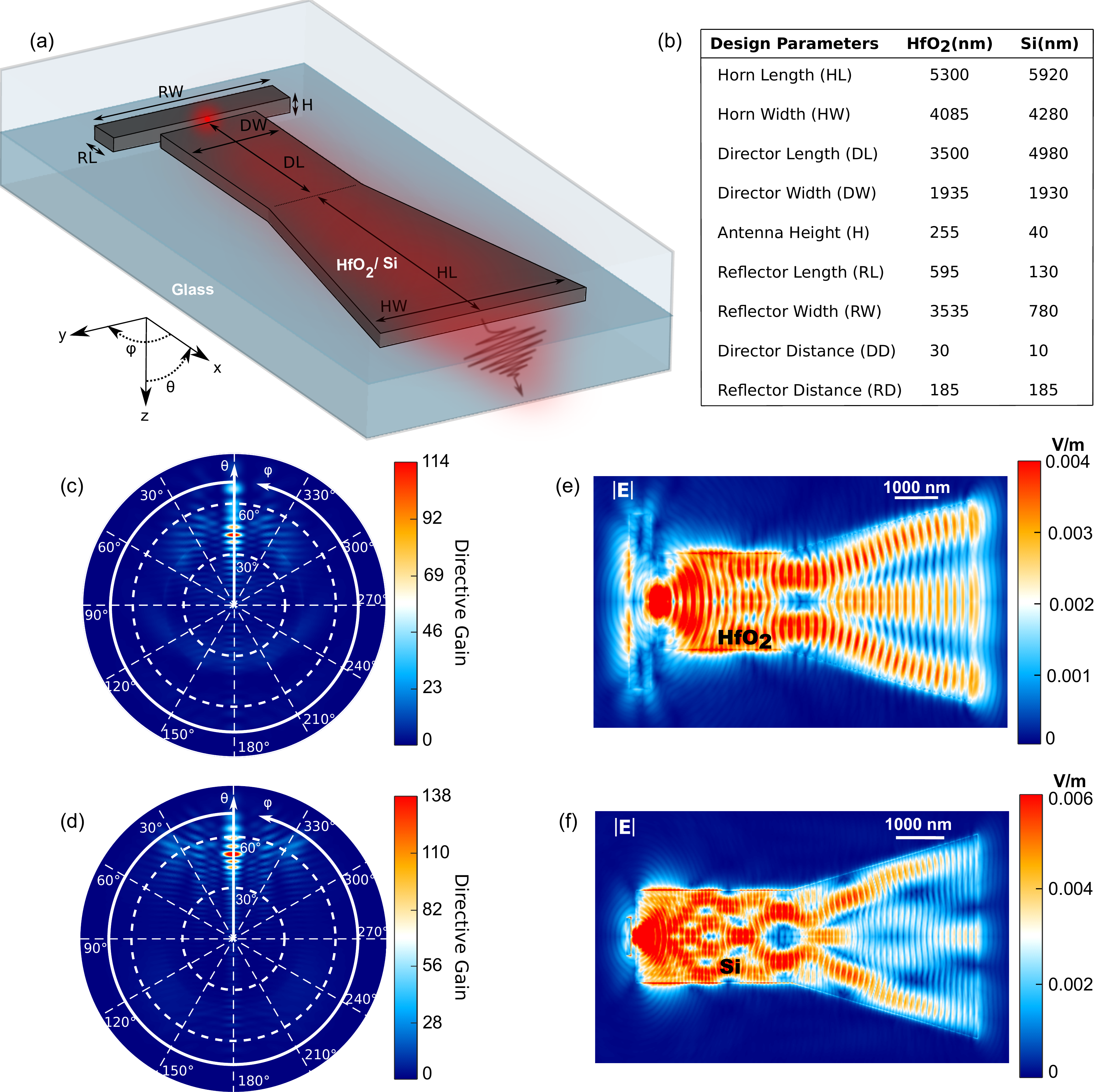}
\caption{Optimization of HfO$_2$ and Si horn-shaped antennas. (a) Schematic representation of our numerical setup. The antenna is composed of the HfO$_2$ or Si horn-shaped director and rectangular-shaped reflector lying on a SiO$_2$ substrate, defined by nine design parameters affecting the directivity: the antenna height (H), director length (DL), director width (DW), horn length (HL), horn width at the flared end (HW), reflector length (RL), reflector width (RW), distance of the field source from the director (DD) and the reflector (RD). The dipole source emitting at $780$\,nm with its dipole moment oriented along $y$-axis is placed into the feed gap between the director and reflector, $10$\,nm above substrate at the origin of $y$-axis (red dot). Light coupled to leaky and guided modes propagates along the director ($x$-axis) and emits radiation into the glass (depicted in red). (b) The table specifies the optimal value of each design parameter for the HfO$_2$ and Si horn-shaped antennas obtained from the optimization process. (c,d) Calculated angular linear directive gain distributions of the optimized (c) HfO$_2$ and (d) Si horn-shaped antennas exhibiting an in-plane directivity of $D=114$ at $\theta=41^{\circ}$ and $D=138$ at $\theta=49^{\circ}$, respectively. (e,f) Calculated absolute value of the electric near-fields $|E|$ (linear scale) of the (e) HfO$_2$ and (f) Si optimized horn antennas in the $xy$-plane at $z=0$, as produced by a $y$-oriented dipole emitter.}
\label{fig::fig4}
\end{figure}
Aiming at further improving the directivity, we extended our study to HfO$_2$ and Si horn-shaped antennas again consisting of a rectangular-shaped reflector and a director which now resembles a H-plane sectoral horn, as illustrated in Fig.~\ref{fig::fig4}a. The optimization setup and design parameters are the same as for the case of rectangular-shaped antennas but we define two additional design parameters for description of the horn-part of the antenna, namely the horn length (HL) and horn width (HW) at the flared end. The optimization process performed for HfO$_2$ and Si horn-shaped antennas results in the geometrical parameters listed in the table of Fig.~\ref{fig::fig4}b. Notably, the optimized horn antennas have a larger footprint than their rectangular-type counterparts, but they also exhibit a narrower radiation pattern with much better directivity. Indeed, the calculated angle-resolved directive gain patterns shown in Fig.~\ref{fig::fig4}c and d reveal tightly focused radiation spots of the main lobe at $\theta=41^{\circ}$ and $\theta=49^{\circ}$ with a directivity of $114$ and $138$ for the HfO$_2$ and Si horn-shaped antennas, respectively. Though these directivities are much better than those for their rectangular-shaped counterparts, both radiation patterns suffer from slightly higher side lobe levels. However, this can be tackled by slight manipulation of the length and aperture of the horn which strongly influence the radiation pattern (not shown here). Interestingly, in spite of different dimensions of the antennas, the angle of their flare is the same ($\approx11^{\circ}$), suggesting its key role in governing the radiation properties including the gain and directivity. Mode analysis for the rectangular section of the director, not shown here, reveals the presence of three TE, two TM and one hybrid mode in the HfO$_2$ antenna, while the antenna composed of Si supports seven TE modes owing to its much smaller height. More higher order guided modes are accommodated in the flared section of the antenna, however, they are not strongly excited. Nevertheless, only two of these guided modes, TE$_{0,0}$ and TE$_{0,2}$ are efficiently excited in both antennas and majorly contribute to the resulting near-field and radiation patterns. In particular, the simulated electric near-fields in Fig.~\ref{fig::fig4}e and f show the interplay of these modes along the director with similar radial-like patterns emerging from the horn section. The negligible back reflections into the horn implicate good impedance matching, which is a valuable feature of the horn antennas. We finally note that, at the operational wavelength, the HfO$_2$ antenna again exhibits a near-unity radiation efficiency, while the Si antenna has a radiation efficiency of $52\%$ owing to its extinction coefficient and the large footprint of the structure.

\subsection{Comparative Study of the Investigated Antennas}
\begin{figure}[h!]
\centering\includegraphics[keepaspectratio,width=\textwidth]{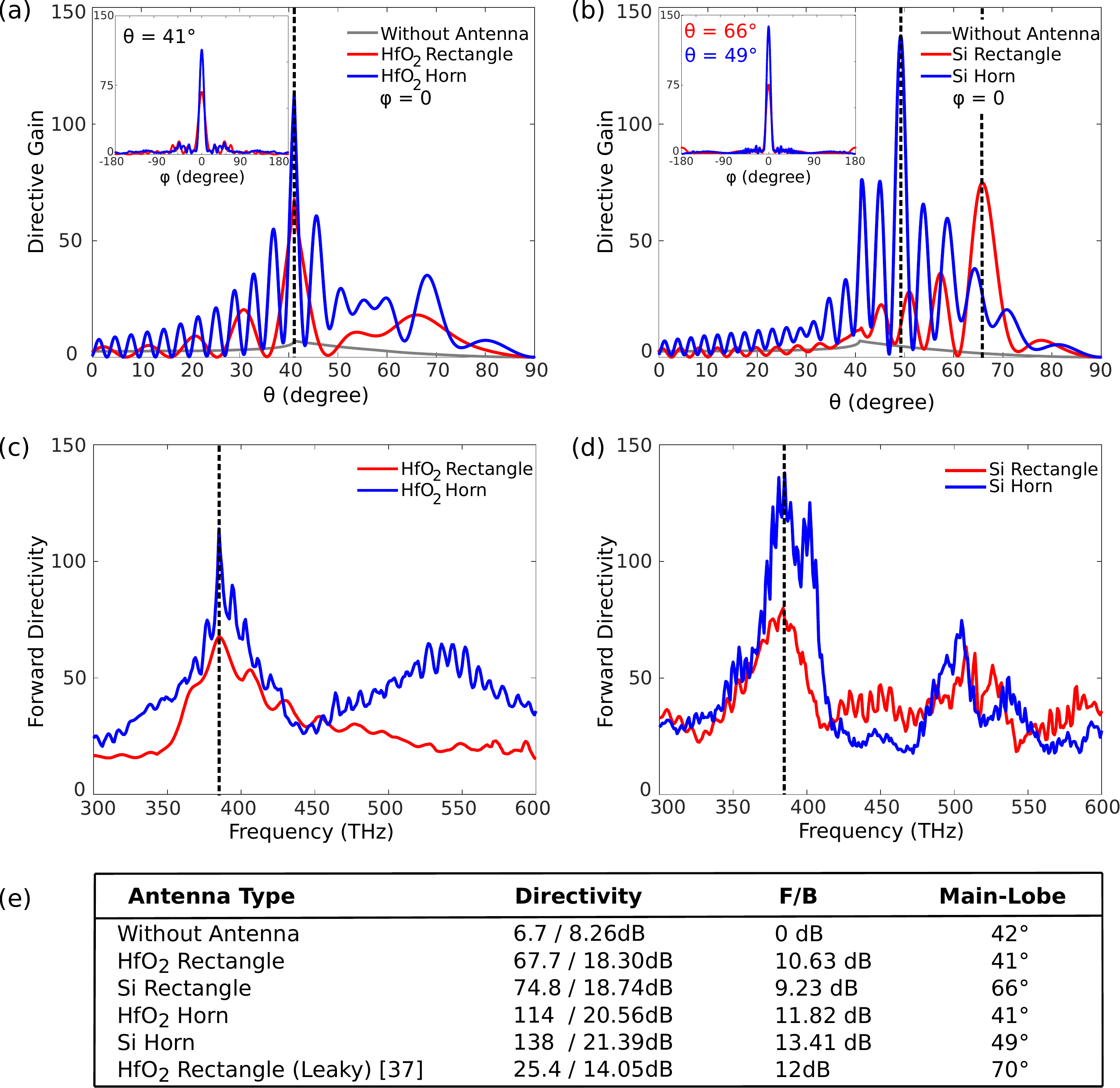}
\caption{Comparison of the characteristics of our four investigated antennas. (a,b) Calculated directive gain of the (a) HfO$_2$ and (b) Si rectangular- (red curve) and horn-shaped (blue curve) antennas as a function of the polar angle $\theta$ at $\varphi=0^\circ$, highlighting the direction of the main lobe and side lobes of the antennas. The grey curves represent the directive gain without an antenna. The insets show the directive gain as a function of the azimuthal angle $\varphi$ at $\theta$ corresponding to the respective main lobe angles, demonstrating in-plane type of propagation. (c,d) Forward directivity of the (c) HfO$_2$ and (d) Si rectangular- (red curve) and horn-shaped (blue curve) antennas as a function of the operational frequency showing the broadband nature of our antennas. The dashed line on each plot indicates the frequency at which the optimization is carried out. (e) The table summarizes the radiation properties such as directivity, angle of the main lobe, front-to-back ratio (F/B) of the proposed antennas and compares them with the reference leaky-wave HfO$_2$ antenna proposed in Ref.~\cite{peter2017directional}.}
\label{fig::fig5}
\end{figure}
To compare the features already suggested by the figures above, in Figure~\ref{fig::fig5}, we summarize the important radiation characteristics of our four investigated antennas. We begin with a comparison of their directive gain plotted as a function of the polar angle $\theta$ in forward direction at $\varphi=0^\circ$, as shown in Figure~\ref{fig::fig5}a and b. The horn-shaped antennas demonstrate significantly increased directivity in comparison to their rectangular-shaped counterparts at the price of increased side lobe level and footprint of the structures. Interestingly, while the height of the antenna has a significant influence on the main lobe angle for the HfO$_2$ antennas, in the Si antennas this angle is mainly influenced by the large contrast and extinction coefficient of the refractive index. Therefore, the HfO$_2$ antennas have their main radiation lobes directed along the same direction which can be attributed to the negligible 5nm change in the height of these two antennas and the Si antennas generate their main lobes at remarkably different angles which is a clear consequence of having different antenna heights which are sensitive to the larger refractive index. Also, unlike the HfO$_2$ counterpart it exhibits a lower radiation efficiency which has an impact on the emission pattern and its main lobe direction. However, all antennas possess a predominantly in-plane radiation ($xz$-plane) with their maximum directive gains along the forward direction at the azimuthal angle $\varphi=0^\circ$ for their respective main lobe directions, $\theta$, as depicted in the insets. Furthermore, the forward directivity of our HfO$_2$ and Si antennas as a function of the operational frequency shown in Figure~\ref{fig::fig5}c and d, respectively, exhibit a broadband nature. In particular, the HfO$_2$ rectangular-shaped antenna maintains a directivity of at least $35$ from $360$ to $420$\,THz and the HfO$_2$ horn-shaped antenna along with both the Si antennas maintain this minimum over the whole range from $300-600$\,THz. This makes our structures robust for operational frequency variations. Finally, the table of Figure~\ref{fig::fig5}e summarizes the radiation properties of the proposed antennas and compares them with the reference leaky-wave HfO$_2$ antenna investigated theoretically and experimentally in \cite{peter2017directional}. In particular, the front-to-back ratio (F/B), which is the ratio of the maximum radiation intensities at $\varphi=0^\circ$ and $\varphi=180^\circ$, is generally higher for the Si antennas than for the HfO$_2$ analogues due to the refractive index contrast between the antenna and substrate. Overall, the proposed guided wave antennas demonstrate much better directivity than the referred leaky-wave antenna.

\section{Conclusion}
In conclusion, we have optimized the emission directivity of four directional travelling wave antennas using particle swarm optimization in conjunction with local trust-region optimization. The dielectric antennas of simple rectangular- and horn-shaped designs are composed of HfO$_2$ or Si directors and reflectors placed on silica substrate with dipole emitters as an internal source of excitation. Our optimized HfO$_2$ and Si rectangular-shaped antennas possess highly directional emission with a linear directivity of 68 and 75, respectively, whereas their horn-shaped analogues respectively demonstrate further improved directivities of 114 and 138. However, this comes at the cost of higher side lobe levels and increased size of the corresponding structures. Our numerical analysis reveals that significantly improved directivity of our designed antennas in comparison to that of the leaky mode antenna originates from the presence of two dominant guided TE modes in the director which couple strongly to the dipole emitter. Thus, these guided modes dominate the far-field emission pattern for each antenna and govern the direction of the main lobe emission which predominately originates from the end facet of the director. Furthermore, our parametric study conducted on all design parameters exposes their respective influence on the directivity and offers a way to tailor desired emission characteristics. We anticipate that the proposed nanoantennas can be experimentally implemented by using the conventional two-step electron beam lithography and pave its way to applications in sensing, optical manipulation, wireless on-chip communications and optical interconnects.

\section*{Funding/Acknowledgment}
H.F., V.M., and J.F. gratefully acknowledge financial support from the Deutsche Forschungsgemeinschaft (DFG) via TRR142 project C05, and computing time support provided by the Paderborn Center for Parallel Computing (PC$^2$).
S.L. and T.L. acknowledge financial support by the German Federal Ministry of Education and Research through the funding program Photonics Research Germany (Project 13N14150) and by the Deutsche Forschungsgemeinschaft, Cluster of Excellence ML4Q (EXC 2004/1–390534769).

\bibliography{main}

\begin{thebibliography}{10}

\bibitem{bharadwaj2009optical}
P.~Bharadwaj, B.~Deutsch, and L.~Novotny, ``Optical antennas,'' {\em Advances
  in Optics and Photonics}, vol.~1, no.~3, pp.~438--483, 2009.

\bibitem{novotny2011antennas}
L.~Novotny and N.~Van~Hulst, ``Antennas for light,'' {\em Nature Photonics},
  vol.~5, no.~2, pp.~83--90, 2011.

\bibitem{bryant2008mapping}
G.~W. Bryant, F.~J. Garc{\'\i}a~de Abajo, and J.~Aizpurua, ``Mapping the
  plasmon resonances of metallic nanoantennas,'' {\em Nano Letters}, vol.~8,
  no.~2, pp.~631--636, 2008.

\bibitem{han2016high}
S.~Han, H.~Kim, Y.~W. Kim, Y.-J. Kim, S.~Kim, I.-Y. Park, and S.-W. Kim,
  ``High-harmonic generation by field enhanced femtosecond pulses in
  metal-sapphire nanostructure,'' {\em Nature Communications}, vol.~7, no.~1,
  pp.~1--7, 2016.

\bibitem{gigli2019quasinormal}
C.~Gigli, T.~Wu, G.~Marino, A.~Borne, G.~Leo, and P.~Lalanne,
  ``Quasinormal-mode modeling and design in nonlinear nano-optics,'' {\em arXiv
  Preprint arXiv:1911.06373}, 2019.

\bibitem{curto2010unidirectional}
A.~G. Curto, G.~Volpe, T.~H. Taminiau, M.~P. Kreuzer, R.~Quidant, and N.~F. van
  Hulst, ``Unidirectional emission of a quantum dot coupled to a nanoantenna,''
  {\em Science}, vol.~329, no.~5994, pp.~930--933, 2010.

\bibitem{li2007shaping}
J.~Li, A.~Salandrino, and N.~Engheta, ``Shaping light beams in the nanometer
  scale: A {Y}agi-{U}da nanoantenna in the optical domain,'' {\em Physical
  Review B}, vol.~76, no.~24, p.~245403, 2007.

\bibitem{krasnok2013optical}
A.~E. Krasnok, I.~S. Maksymov, A.~I. Denisyuk, P.~A. Belov, A.~E.
  Miroshnichenko, C.~R. Simovski, and Y.~S. Kivshar, ``Optical nanoantennas,''
  {\em Physics-Uspekhi}, vol.~56, no.~6, p.~539, 2013.

\bibitem{giannini2011plasmonic}
V.~Giannini, A.~I. Fern{\'a}ndez-Dom{\'\i}nguez, S.~C. Heck, and S.~A. Maier,
  ``Plasmonic nanoantennas: Fundamentals and their use in controlling the
  radiative properties of nanoemitters,'' {\em Chemical Reviews}, vol.~111,
  no.~6, pp.~3888--3912, 2011.

\bibitem{halas2011plasmons}
N.~J. Halas, S.~Lal, W.-S. Chang, S.~Link, and P.~Nordlander, ``Plasmons in
  strongly coupled metallic nanostructures,'' {\em Chemical Reviews}, vol.~111,
  no.~6, pp.~3913--3961, 2011.

\bibitem{hoang2015ultrafast}
T.~B. Hoang, G.~M. Akselrod, C.~Argyropoulos, J.~Huang, D.~R. Smith, and M.~H.
  Mikkelsen, ``Ultrafast spontaneous emission source using plasmonic
  nanoantennas,'' {\em Nature Communications}, vol.~6, no.~1, pp.~1--7, 2015.

\bibitem{mongia1994dielectric}
R.~K. Mongia and P.~Bhartia, ``Dielectric resonator antennas—{A} review and
  general design relations for resonant frequency and bandwidth,'' {\em
  International Journal of Microwave and Millimeter-Wave Computer-Aided
  Engineering}, vol.~4, no.~3, pp.~230--247, 1994.

\bibitem{kuznetsov2016optically}
A.~I. Kuznetsov, A.~E. Miroshnichenko, M.~L. Brongersma, Y.~S. Kivshar, and
  B.~Luk’yanchuk, ``Optically resonant dielectric nanostructures,'' {\em
  Science}, vol.~354, no.~6314, p.~aag2472, 2016.

\bibitem{au2019high}
T.~H. Au, S.~Buil, X.~Qu{\'e}lin, J.-P. Hermier, and N.~D. Lai, ``High
  directional radiation of single photon emission in a dielectric antenna,''
  {\em ACS Photonics}, vol.~6, no.~11, pp.~3024--3031, 2019.

\bibitem{krasnok2012all}
A.~E. Krasnok, A.~E. Miroshnichenko, P.~A. Belov, and Y.~S. Kivshar,
  ``All-dielectric optical nanoantennas,'' {\em Optics Express}, vol.~20,
  no.~18, pp.~20599--20604, 2012.

\bibitem{barreda2019recent}
A.~Barreda, J.~Saiz, F.~Gonz{\'a}lez, F.~Moreno, and P.~Albella, ``Recent
  advances in high refractive index dielectric nanoantennas: Basics and
  applications,'' {\em AIP Advances}, vol.~9, no.~4, p.~040701, 2019.

\bibitem{sain2019nonlinear}
B.~Sain, C.~Meier, and T.~Zentgraf, ``Nonlinear optics in all-dielectric
  nanoantennas and metasurfaces: A review,'' {\em Advanced Photonics}, vol.~1,
  no.~2, p.~024002, 2019.

\bibitem{cambiasso2017bridging}
J.~Cambiasso, G.~Grinblat, Y.~Li, A.~Rakovich, E.~Cort{\'e}s, and S.~A. Maier,
  ``Bridging the gap between dielectric nanophotonics and the visible regime
  with effectively lossless {G}allium {P}hosphide antennas,'' {\em Nano
  Letters}, vol.~17, no.~2, pp.~1219--1225, 2017.

\bibitem{bidault2019dielectric}
S.~Bidault, M.~Mivelle, and N.~Bonod, ``Dielectric nanoantennas to manipulate
  solid-state light emission,'' {\em Journal of Applied Physics}, vol.~126,
  no.~9, p.~094104, 2019.

\bibitem{morozov2018metal}
S.~Morozov, M.~Gaio, S.~A. Maier, and R.~Sapienza, ``Metal--dielectric
  parabolic antenna for directing single photons,'' {\em Nano Letters},
  vol.~18, no.~5, pp.~3060--3065, 2018.

\bibitem{livneh2016highly}
N.~Livneh, M.~G. Harats, D.~Istrati, H.~S. Eisenberg, and R.~Rapaport, ``Highly
  directional room-temperature single photon device,'' {\em Nano Letters},
  vol.~16, no.~4, pp.~2527--2532, 2016.

\bibitem{rusak2014hybrid}
E.~Rusak, I.~Staude, M.~Decker, J.~Sautter, A.~E. Miroshnichenko, D.~A. Powell,
  D.~N. Neshev, and Y.~S. Kivshar, ``Hybrid nanoantennas for directional
  emission enhancement,'' {\em Applied Physics Letters}, vol.~105, no.~22,
  p.~221109, 2014.

\bibitem{biagioni2012nanoantennas}
P.~Biagioni, J.-S. Huang, and B.~Hecht, ``Nanoantennas for visible and infrared
  radiation,'' {\em Reports on Progress in Physics}, vol.~75, no.~2, p.~024402,
  2012.

\bibitem{TWA}
R.~Elliott, {\em Traveling Wave Antennas}.
\newblock McGraw-Hill, 1965.

\bibitem{constantine2005antenna}
C.~A. Balanis, {\em Antenna theory: Analysis and Design}.
\newblock Wiley-Interscience, 1982.

\bibitem{Antenny}
G.~Z. Aizenberg, {\em Antenny ul\'trakorotkikh voln [part 1]}.
\newblock Moscow, 1957.

\bibitem{agio2013optical}
M.~Agio and A.~Al{\`u}, {\em Optical antennas}.
\newblock Cambridge University Press, 2013.

\bibitem{li2021directional}
N.~Li, Y.~Lai, S.~H. Lam, H.~Bai, L.~Shao, and J.~Wang, ``Directional control
  of light with nanoantennas,'' {\em Advanced Optical Materials}, vol.~9,
  no.~1, p.~2001081, 2021.

\bibitem{milligan2005modern}
T.~A. Milligan, {\em Modern antenna design}.
\newblock John Wiley \& Sons, 2005.

\bibitem{sutinjo2008radiation}
A.~Sutinjo, M.~Okoniewski, and R.~H. Johnston, ``Radiation from fast and slow
  traveling waves,'' {\em IEEE Antennas and Propagation Magazine}, vol.~50,
  no.~4, pp.~175--181, 2008.

\bibitem{oliner2007leaky}
A.~A. Oliner, D.~R. Jackson, and J.~Volakis, ``Leaky-wave antennas,'' {\em
  Antenna Engineering Handbook}, vol.~4, p.~12, 2007.

\bibitem{hu2009understanding}
J.~Hu and C.~R. Menyuk, ``Understanding leaky modes: Slab waveguide
  revisited,'' {\em Advances in Optics and Photonics}, vol.~1, no.~1,
  pp.~58--106, 2009.

\bibitem{peng1981guidance}
S.-T. Peng and A.~A. Oliner, ``Guidance and leakage properties of a class of
  open dielectric waveguides: Part i-{M}athematical formulations,'' {\em IEEE
  Transactions on Microwave Theory and Techniques}, vol.~29, no.~9,
  pp.~843--855, 1981.

\bibitem{oliner1981guidance}
A.~Oliner, S.-T. Peng, T.-I. Hsu, and A.~Sanchez, ``Guidance and leakage
  properties of a class of open dielectric waveguides: Part ii-{N}ew physical
  effects,'' {\em IEEE Transactions on Microwave Theory and Techniques},
  vol.~29, no.~9, pp.~855--869, 1981.

\bibitem{torner1990leaky}
L.~Torner, F.~Canal, and J.~Hernandez-Marco, ``Leaky modes in multilayer
  uniaxial optical waveguides,'' {\em Applied Optics}, vol.~29, no.~18,
  pp.~2805--2814, 1990.

\bibitem{tamir1986varieties}
T.~Tamir and F.~Kou, ``Varieties of leaky waves and their excitation along
  multilayered structures,'' {\em IEEE Journal of Quantum Electronics},
  vol.~22, no.~4, pp.~544--551, 1986.

\bibitem{peter2017directional}
M.~Peter, A.~Hildebrandt, C.~Schlickriede, K.~Gharib, T.~Zentgraf,
  J.~F\"orstner, and S.~Linden, ``Directional emission from dielectric
  leaky-wave nanoantennas,'' {\em Nano Letters}, vol.~17, no.~7,
  pp.~4178--4183, 2017.

\bibitem{leuteritz2021dielectric}
T.~Leuteritz, H.~Farheen, S.~Qiao, F.~Spreyer, C.~Schlickriede, T.~Zentgraf,
  V.~Myroshnychenko, J.~F\"orstner, and S.~Linden, ``Dielectric travelling wave
  antennas for directional light emission,'' {\em Optics Express}, vol.~29,
  no.~10, pp.~14694--14704, 2021.

\bibitem{hansen1940radiating}
W.~Hansen, ``Radiating electromagnetic waveguide,'' {\em US Patent A},
  vol.~2402622, 1940.

\bibitem{jackson2012leaky}
D.~R. Jackson, C.~Caloz, and T.~Itoh, ``Leaky-wave antennas,'' {\em Proceedings
  of the IEEE}, vol.~100, no.~7, pp.~2194--2206, 2012.

\bibitem{mohsen2018fundamental}
M.~K. Mohsen, M.~M. Isa, A.~Isa, M.~Zin, S.~Saat, Z.~Zakaria, I.~Ibrahim,
  M.~Abu, A.~Ahmad, and M.~Abdulhameed, ``The fundamental of leaky wave
  antenna,'' {\em Journal of Telecommunication, Electronic and Computer
  Engineering (JTEC)}, vol.~10, no.~1, pp.~119--127, 2018.

\bibitem{jackson2013recent}
D.~R. Jackson, ``Recent advances in leaky-wave antennas,'' in {\em 2013
  International Symposium on Electromagnetic Theory}, pp.~9--12, IEEE, 2013.

\bibitem{jackson2008role}
D.~Jackson, J.~Chen, R.~Qiang, F.~Capolino, and A.~Oliner, ``The role of leaky
  plasmon waves in the directive beaming of light through a subwavelength
  aperture,'' {\em Optics Express}, vol.~16, no.~26, pp.~21271--21281, 2008.

\bibitem{song2011silicon}
Q.~Song, S.~Campione, O.~Boyraz, and F.~Capolino, ``Silicon-based optical leaky
  wave antenna with narrow beam radiation,'' {\em Optics Express}, vol.~19,
  no.~9, pp.~8735--8749, 2011.

\bibitem{yousefi2012waveguide}
L.~Yousefi and A.~C. Foster, ``Waveguide-fed optical hybrid plasmonic patch
  nano-antenna,'' {\em Optics Express}, vol.~20, no.~16, pp.~18326--18335,
  2012.

\bibitem{molesky2018inverse}
S.~Molesky, Z.~Lin, A.~Y. Piggott, W.~Jin, J.~Vuckovi{\'c}, and A.~W.
  Rodriguez, ``Inverse design in nanophotonics,'' {\em Nature Photonics},
  vol.~12, no.~11, pp.~659--670, 2018.

\bibitem{ma2020deep}
W.~Ma, Z.~Liu, Z.~A. Kudyshev, A.~Boltasseva, W.~Cai, and Y.~Liu, ``Deep
  learning for the design of photonic structures,'' {\em Nature Photonics},
  pp.~1--14, 2020.

\bibitem{jiang2020deep}
J.~Jiang, M.~Chen, and J.~A. Fan, ``Deep neural networks for the evaluation and
  design of photonic devices,'' {\em Nature Reviews Materials}, pp.~1--22,
  2020.

\bibitem{wiecha2019design}
P.~R. Wiecha, C.~Majorel, C.~Girard, A.~Cuche, V.~Paillard, O.~L. Muskens, and
  A.~Arbouet, ``Design of plasmonic directional antennas via evolutionary
  optimization,'' {\em Optics Express}, vol.~27, no.~20, pp.~29069--29081,
  2019.

\bibitem{feichtner2012evolutionary}
T.~Feichtner, O.~Selig, M.~Kiunke, and B.~Hecht, ``Evolutionary optimization of
  optical antennas,'' {\em Physical Review Letters}, vol.~109, no.~12,
  p.~127701, 2012.

\bibitem{briones2018particle}
E.~Briones, R.~Ruiz-Cruz, J.~Briones, N.~Gonzalez, J.~Simon, M.~Arreola, and
  G.~Alvarez-Alvarez, ``Particle swarm optimization of nanoantenna-based
  infrared detectors,'' {\em Optics Express}, vol.~26, no.~22,
  pp.~28484--28496, 2018.

\bibitem{robinson2002particle}
J.~Robinson, S.~Sinton, and Y.~Rahmat-Samii, ``Particle swarm, genetic
  algorithm, and their hybrids: Optimization of a profiled corrugated horn
  antenna,'' in {\em IEEE Antennas and Propagation Society International
  Symposium (IEEE Cat. No. 02CH37313)}, vol.~1, pp.~314--317, IEEE, 2002.

\bibitem{CST}
Dassault Syst{\'e}mes, "CST Studio Suite,"
  \href{https://www.cst.com}{https://www.cst.com}.

\bibitem{kennedy1995particle}
J.~Kennedy and R.~Eberhart, ``Particle swarm optimization,'' in {\em
  Proceedings of ICNN'95-International Conference on Neural Networks}, vol.~4,
  pp.~1942--1948, IEEE, 1995.

\bibitem{sun2013large}
J.~Sun, E.~Timurdogan, A.~Yaacobi, Z.~Su, E.~S. Hosseini, D.~B. Cole, and M.~R.
  Watts, ``Large-scale {S}ilicon photonic circuits for optical phased arrays,''
  {\em IEEE Journal of Selected Topics in Quantum Electronics}, vol.~20, no.~4,
  pp.~264--278, 2013.

\bibitem{lechago2019all}
S.~Lechago, C.~Garc{\'\i}a-Meca, A.~Griol, M.~Kovylina, L.~Bellieres, and
  J.~Mart{\'\i}, ``All-silicon on-chip optical nanoantennas as efficient
  interfaces for plasmonic devices,'' {\em ACS Photonics}, vol.~6, no.~5,
  pp.~1094--1099, 2019.

\bibitem{staude2017metamaterial}
I.~Staude and J.~Schilling, ``Metamaterial-inspired {S}ilicon nanophotonics,''
  {\em Nature Photonics}, vol.~11, no.~5, pp.~274--284, 2017.

\bibitem{kivshar2018all}
Y.~Kivshar, ``All-dielectric meta-optics and non-linear nanophotonics,'' {\em
  National Science Review}, vol.~5, no.~2, pp.~144--158, 2018.

\end{thebibliography}
\end{document}